\documentclass[letters,fleqn,usenatbib]{mnras}

\usepackage{mathptmx}
\usepackage{txfonts}

\usepackage[T1]{fontenc}
\usepackage{ae,aecompl}

\usepackage{graphicx}	% Including figure files
\usepackage{amssymb}	% Extra maths symbols

\newcommand{\hii}{H\,{\scriptsize II}}

\newcommand{\av}{A$_{\rm V}$}

\title[Detection of two power-law tails in PDFs of massive GMCs]{Detection of two power-law tails in the probability distribution functions of massive GMCs}

\author[N.Schneider et al.]{N. Schneider,$^{1,2}$\thanks{E-mail: nicola.schneider@obs.u-bordeaux1.fr}
S.Bontemps$^{1}$,
P.Girichidis$^{3}$,
T.Rayner$^{4}$,
F.Motte$^{5}$,
Ph.Andr\'e$^{5}$,
\newauthor
D.Russeil$^{6}$,
A.Abergel$^{7}$,
L.Anderson$^{8}$,
D.Arzoumanian$^{7}$,
M.Benedettini$^{9}$,
T.Csengeri$^{10}$,
\newauthor
P.Didelon$^{5}$,
J.Di Francesco$^{11}$,
M.Griffin$^{4}$,
T.Hill$^{12}$,
R.S.Klessen$^{13}$,
V.Ossenkopf$^{2}$,
\newauthor
S.Pezzuto$^{9}$,
A.Rivera-Ingraham$^{14}$,
L.Spinoglio$^{9}$,
P.Tremblin$^{15,5}$, 
A. Zavagno$^{5}$ \\
$^1$OASU/LAB Univ. Bordeaux, CNRS, UMR5804, 33270 Floirac, France\\
$^2$I. Physik. Institut, University of Cologne, 50937 Cologne, Germany\\ 
$^3$Max Planck Institut f\"ur Astrophysik, 85741 Garching, Germany\\ 
$^4$School of Physics and Astronomy, Cardiff University, CF24 3AA, UK\\
$^5$IRFU/SAp CEA/DSM, Lab. AIM CNRS - Universit\'e Paris Diderot, 91191 Gif-sur-Yvette, France\\
$^6$AIX Marseille Univ. LAM, CNRS, UMR 7326, 13388 Marseille, France\\
$^7$IAS, CNRS, UMR8617, Universit\'e Paris Sud, 91400 Orsay, France\\
$^8$Department of Physics and Astronomy, West Virginia University, WV 26506, USA\\
$^9$IAPS, INAF, 00133 Roma, Italy\\
$^{10}$Max Planck Institut f\"ur Radioastronomie, 53121 Bonn, Germany\\
$^{11}$NRC, Herzberg Institute of Astrophysics, Victoria, Canada\\ 
$^{12}$Joint ALMA Observatory, Santiago, Chile\\
$^{13}$Universit\"at Heidelberg, Zentrum f\"ur Astronomie, 69120  Heidelberg, Germany\\ 
$^{14}$ESA/ESAC, Madrid, Spain\\
$^{15}$Astrophysics Group, University of Exeter, UK\\
}

\date{Accepted 2015 July 21. Received 2015 July 20; in original form 2015 June 29}

\pubyear{2015}

\begin{document}
\label{firstpage}
\pagerange{\pageref{firstpage}--\pageref{lastpage}}
\maketitle

\begin{abstract}
We report the novel detection of complex high-column density tails in
the probability distribution functions (PDFs) for three high-mass
star-forming regions (CepOB3, MonR2, NGC6334), obtained from dust
emission observed with {\sl Herschel}.  The low column density range
can be fit with a lognormal distribution.  A first power-law tail
starts above an extinction (\av) of $\sim$6--14. It has a slope of
$\alpha$=1.3--2 for the $\rho\propto r^{-\alpha}$ profile for an
equivalent density distribution (spherical or cylindrical geometry),
and is thus consistent with free-fall gravitational collapse. Above
\av $\sim$40, 60, and 140, we detect an excess that can be fitted by a
flatter power law tail with $\alpha>$2. It correlates with the central
regions of the cloud (ridges/hubs) of size $\sim$1 pc and densities
above 10$^4$ cm$^{-3}$. This excess may be caused by physical
processes that slow down collapse and reduce the flow of mass towards
higher densities. Possible are: 1. rotation, which introduces an
angular momentum barrier, 2. increasing optical depth and weaker
cooling, 3. magnetic fields, 4. geometrical effects, and
5. protostellar feedback.  The excess/second power-law tail is closely
linked to high-mass star-formation though it does not imply a
universal column density threshold for the formation of (high-mass)
stars.
\end{abstract}


%%%%%%%%%%%%%%%%%%%%%%%%%%%%%%%%%%%%%%%%%%%%%%%%%%

%%%%%%%%%%%%%%%%% BODY OF PAPER %%%%%%%%%%%%%%%%%%

\section{Introduction} \label{intro}
Probability distribution functions (PDFs) of column density $N$(HI+H$_2$) 
obtained using far-infrared emission of dust from {\sl
  Herschel}\footnote{{\sl Herschel} is an ESA space observatory with
  science instruments provided by European-led Principal Investigator
  consortia and with important participation from NASA.}
show a characteristic shape for low-mass star-forming regions: a
lognormal distribution for low $N$, commonly attributed to turbulence,
and a single power-law tail for higher $N$ \citep[e.g.,][and
  references therein]{schneider2013}.  This sort of PDF is also found
for extinction maps \citep{kai2009,froebrich2010}.
There is an ongoing discussion whether this power-law tail can be
attributed to self-gravity 
\citep{klessen2000,kritsuk2011,fed2013,giri2014,schneider2013,schneider2015a}
or is pressure-driven \citep{kai2011}.

In this {\sl Letter}, we show that PDFs obtained from {\sl Herschel}
dust column density maps of giant molecular clouds (GMCs) not only
show a clear lognormal plus power-law tail distribution but can also
exhibit an excess at very high column densities that can be identified
as a {\sl second}, shallower power law.  It seems that this discovery
is restricted to the densest regions of massive molecular clouds
(ridges/hubs), though not {\sl all} those GMCs display this feature.
The objective of this study is to report our detection and give some
tentative explanations for its existance.

%\begin{table*}[ht]  \label{table}
\begin{table*} \label{table}
\begin{center}
  \caption{Properties of GMCs: total mass $M$
    derived above the first closed contour, i.e. \av=2 for CepOB3 and
    MonR2 and \av=4 for NGC6334, masses (the percentage of the total
    mass given in parenthesis) above the first (DP1) and second (DP2)
    deviation point in the PDF (values determined from the {\sl
      Herschel} N-maps), average and maximum UV-flux in units of the
    Habing field (obtained from the {\sl Herschel} fluxes \citep{rocca2013}), slopes $s1/s2$ of
    power-law tails, exponents $\alpha_{s1/2}$ and $\alpha_{c1/2}$ for
    a spherical and cylindrical density distribution, equivalent beam
    deconvolved radius $r$ (taking an area $A$=$\pi r^2$), and density
    $n$, determined from the average column density and
    $r$ with $n = N/(2r)$.  The values of $r$ and $n$ for NGC6334 are
    averages over several clumps. The errors of
    both $\alpha_c$ and $\alpha_s$ determined from the first slope are
    $\sim$0.02, and those from the second slope are $\sim$0.05.}
\begin{tabular}{lccccccccccccc}
              &                      &             &       &       &    & &         &            &    &  & \\
              &  {\small $M_{total}$} & {\small $M(DP1)$} & {\small $M(DP2)$} & $\langle UV \rangle$ & UV$_{max}$ & $s_1$ & $s_2$ & $\alpha_{s1}$ & $\alpha_{c1}$ & $\alpha_{s2}$ & $\alpha_{c2}$ & r & $\langle n \rangle$ \\
              & {\tiny [10$^4$ M$_{\odot}$]} & {\tiny [10$^4$ M$_{\odot}$]} & {\tiny [10$^4$ M$_{\odot}$]} & {\tiny [G$_o$]} & {\tiny [10$^4$G$_o$]} & & & & & & & 
{\tiny [pc]} &  {\tiny [10$^4$ cm$^{-3}$]} \\
\hline
{\small {\sl CepOB3}} & 10.0 & 1.7 {\tiny(17\%)} & 0.052 {\tiny(0.5\%)}&  40 & 8.0 & -3.80 & -1.18 & 1.53 & 1.26 & 2.70 & 1.85 & 0.29 & 6.0 \\  
{\small {\sl MonR2}}   &  1.4 & 0.54 {\tiny(39\%)}& 0.21 {\tiny(15\%)} &  52 & 4.8 & -2.10 & -1.05 & 1.95 & 1.48 & 2.90 & 1.95 & 0.65 & 1.9 \\ 
{\small {\sl NGC6334}}&  23.1 & 7.3 {\tiny(32\%)} & 3.53 {\tiny(15\%)} & 302 & 8.9 & -2.26 & -0.61 & 1.88 & 1.44 & 4.17 & 2.64 & 0.42 & 34.7 \\  
\hline
\end{tabular}
\end{center}
\end{table*}

\begin{figure*} 
\begin{center}
\includegraphics[angle=0,totalheight=4.8cm]{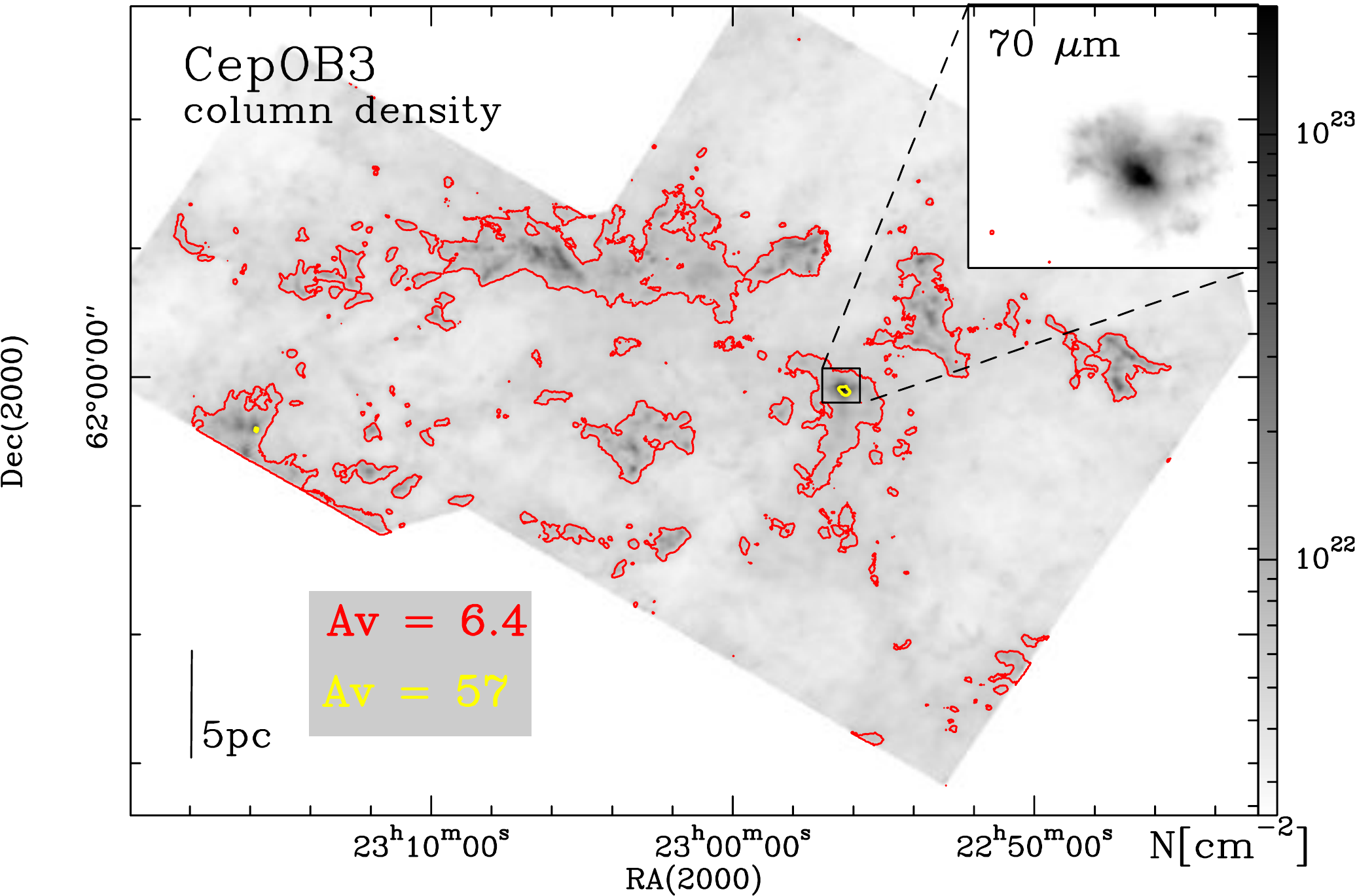}
\hspace{0.2cm}\includegraphics[angle=0,totalheight=5.2cm]{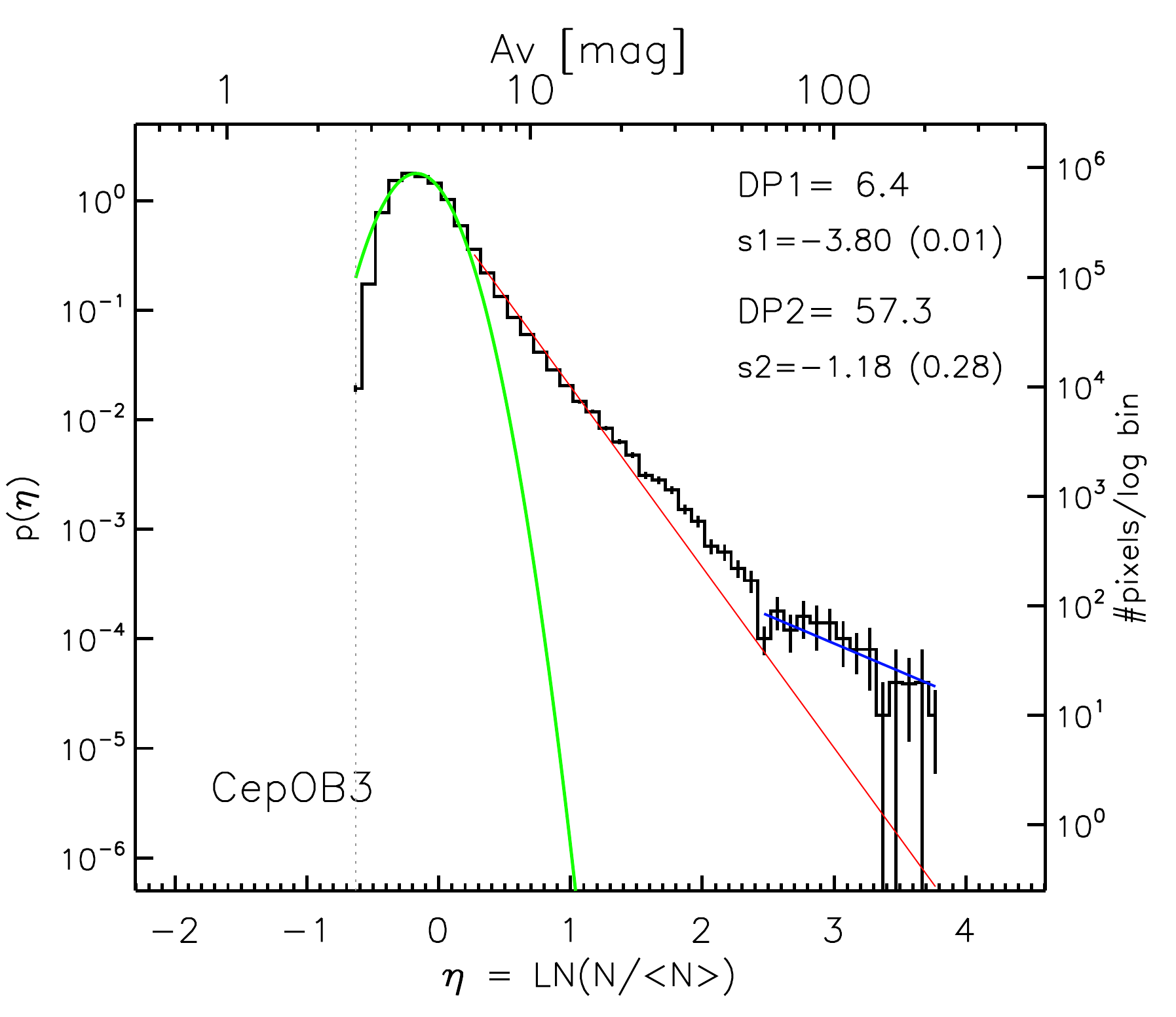}

\vspace{0.4cm}\includegraphics[angle=0,totalheight=5.8cm]{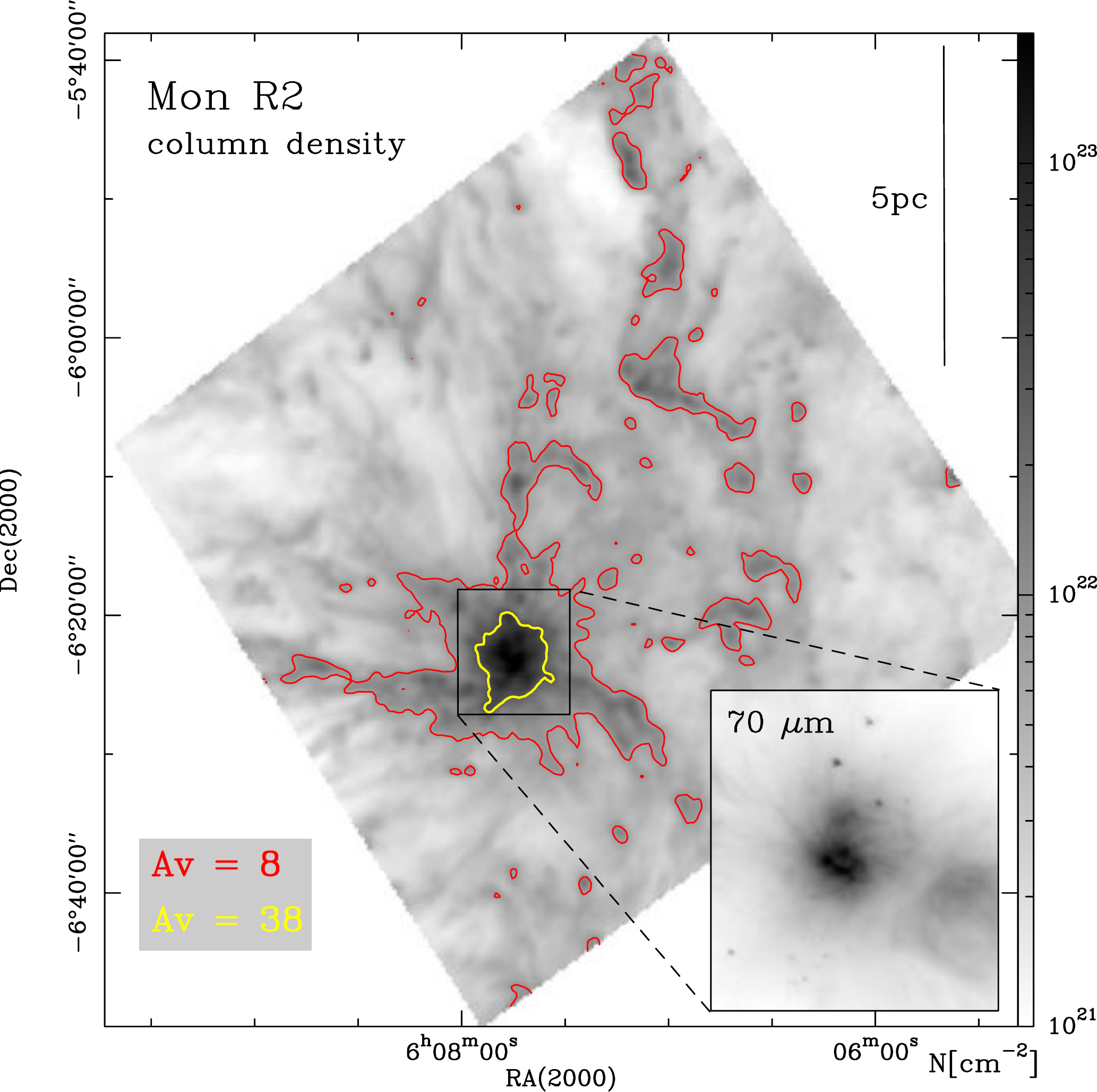}
\hspace{0.8cm}\includegraphics[angle=0,totalheight=5.5cm]{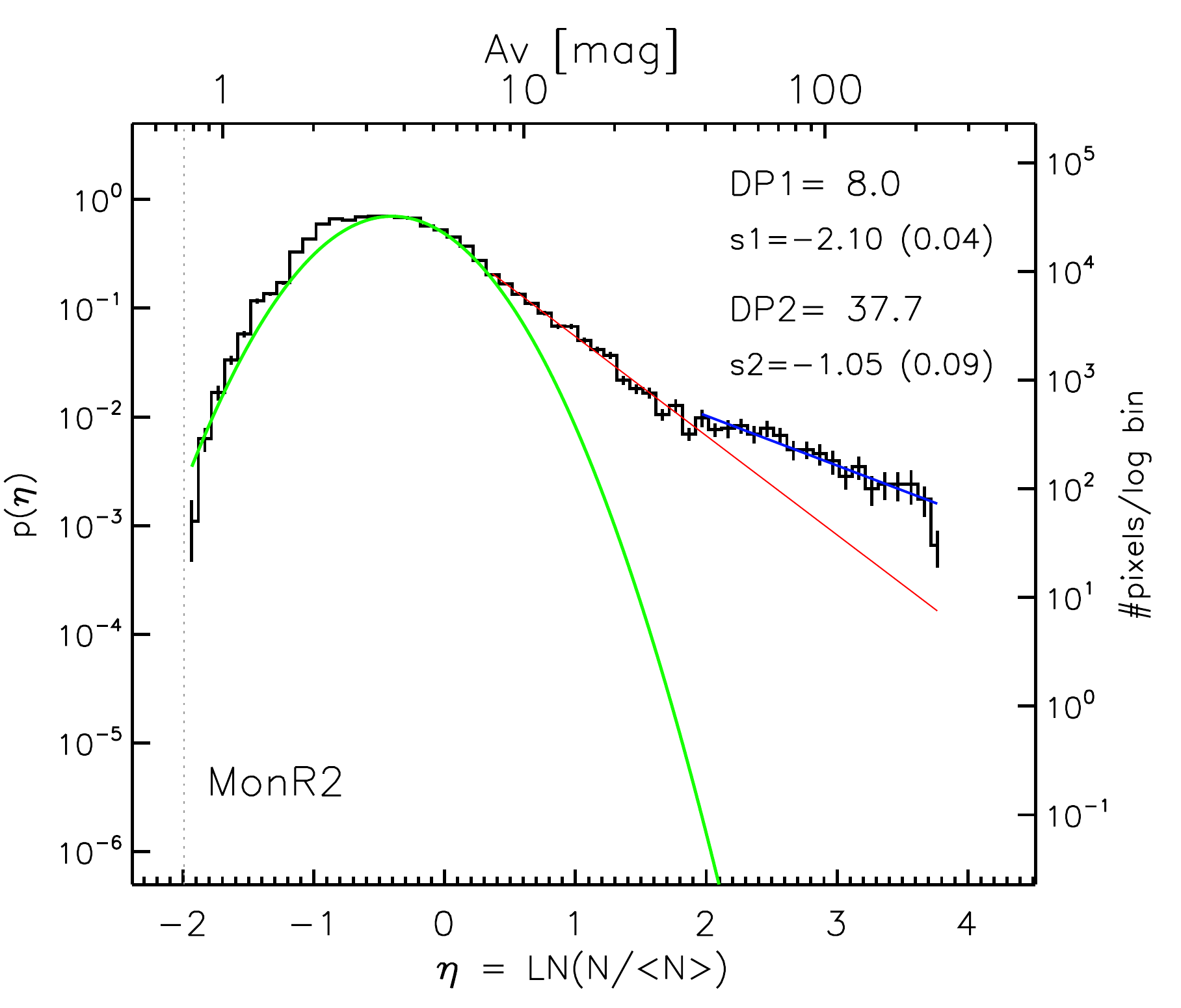}

\vspace{0.4cm}\includegraphics[angle=0,totalheight=5.8cm]{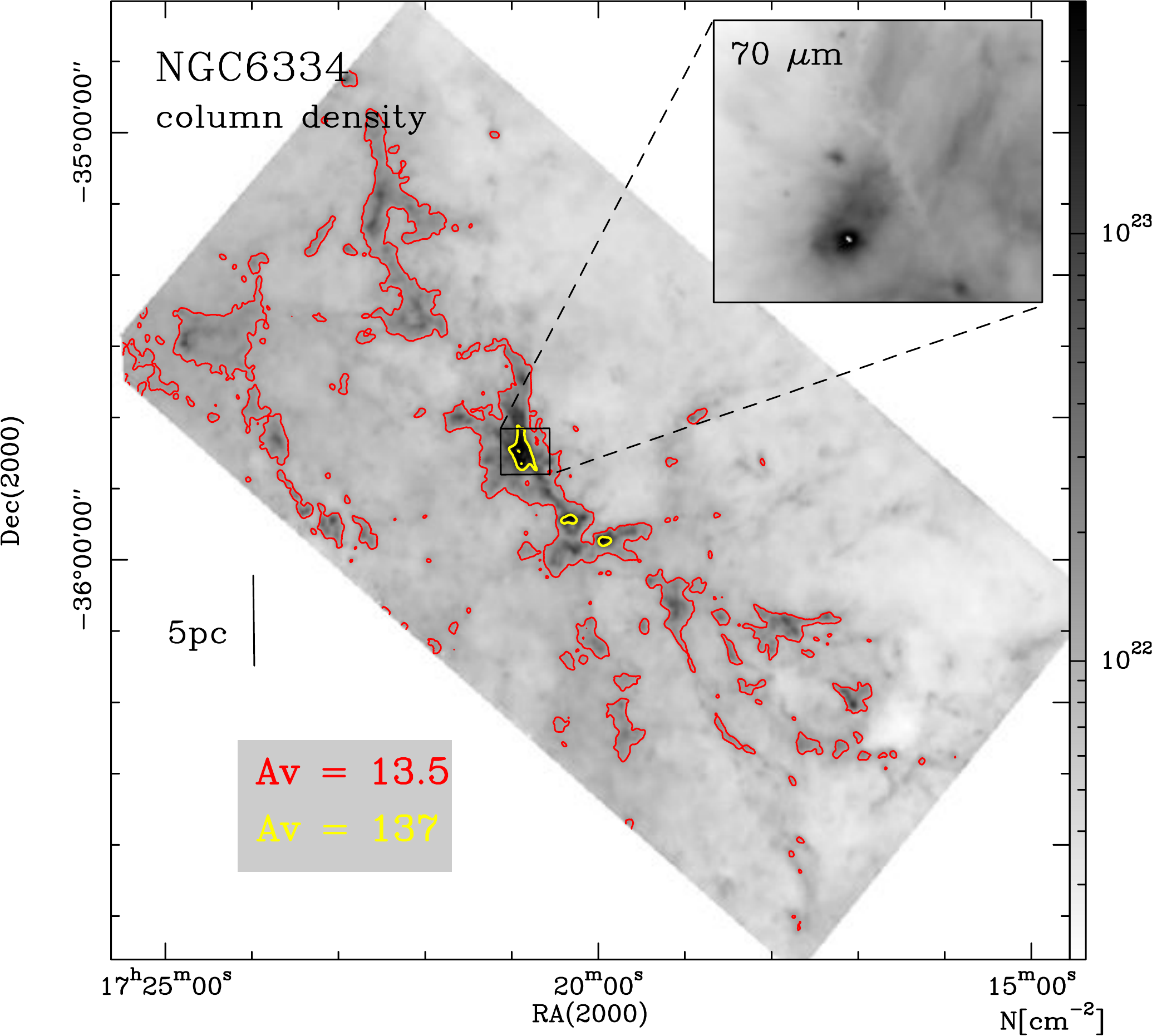}
\hspace{0.8cm}\includegraphics[angle=0,totalheight=5.5cm]{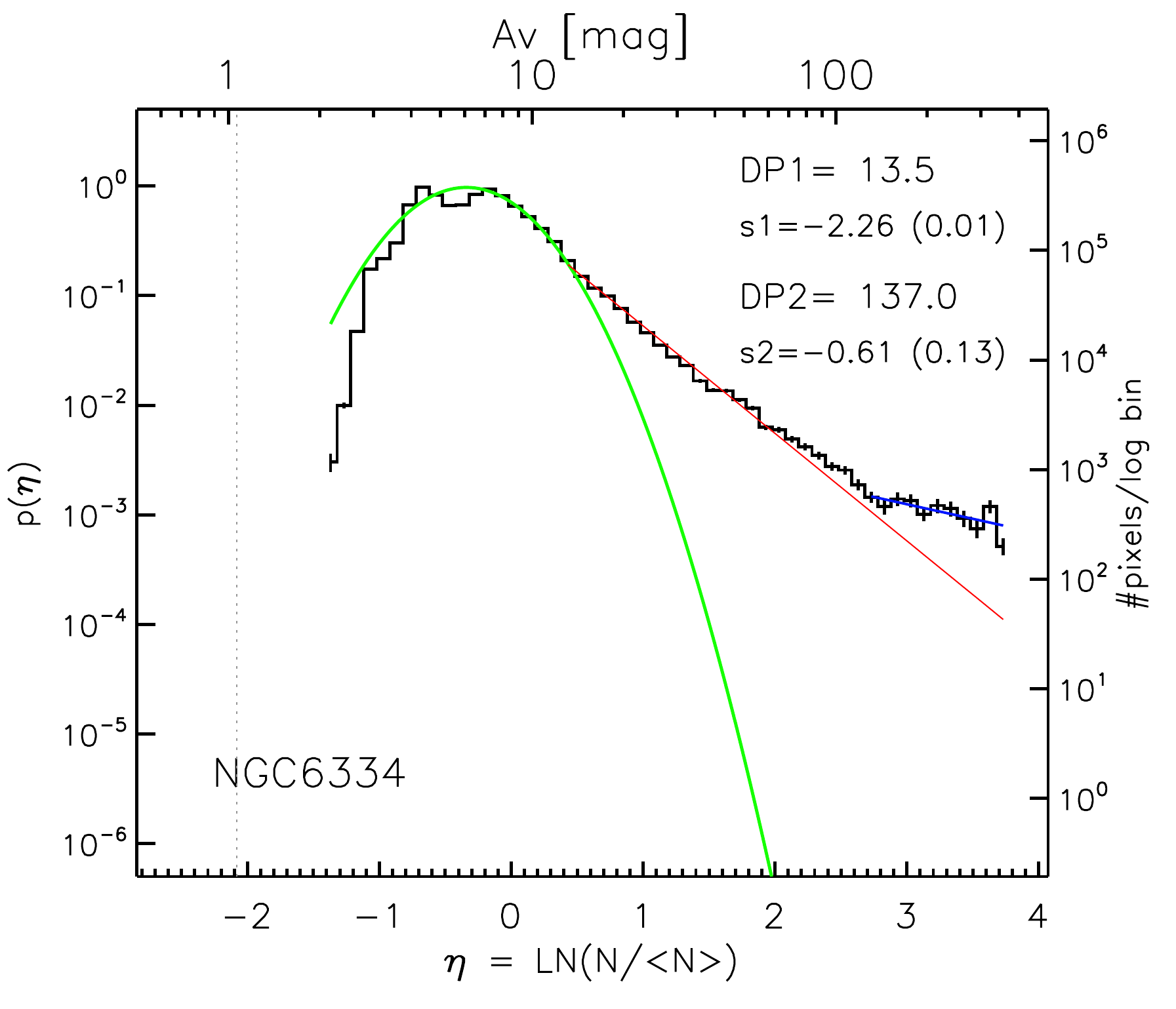}

\end{center}
\caption [] {{\bf Left:} Dust column density maps of the GMCs.
  Contour lines in red and yellow indicate the break points between the 
  lognormal distribution and the first and second power-law tail,
  respectively. The small panels show a zoom into the central regions, displaying
  PACS 70 $\mu$m emission (scale typically 0.1-30 Jy/pixel).
  {\bf Right:} Corresponding PDFs at an angular
  resolution of 36$''$ (binsize 0.1).  The left y-axis indicates the
  normalized probability $p(\eta)$, the right y-axis the number of
  pixels per logarithmic bin.  The error-bars were calculated using
  Poisson-statistics.  The upper x-axis gives the extinction \av\, and
  the lower x-axis $\eta=\ln(N/\langle N \rangle)$. The green curve
  indicates the fitted lognormal PDF and the red and blue lines the
  power law fits to the high-density tails. The fit to the first tail
  was only performed between the break points but the line was
  continued to emphasize the excess.  The deviation points
  (DP) and slopes ($s$ with its error) of the power-law tails are given in each
  panel. }
\label{pdf1} 
\end{figure*}

\vspace{-0.6cm}
\section{Observations} \label{obs}

We make use of {\sl Herschel}-derived column density ($N$) PDFs of
three well-known GMCs. 
%that all have a mass larger than
%10$^4$ M$_\odot$, and form actively massive stars (Sec.~\ref{areas}).
Data and PDFs of MonR2 at a distance of 0.83 kpc (Rayner et al., in
prep, Didelon et al. 2015) and NGC6334 at a distance of 1.35 kpc
\citep{russeil2013} were obtained within the HOBYS ({\it Herschel}
imaging survey of OB Young Stellar objects) key program
\citep{motte2010}, while the data for CepOB3 (distance 0.7 kpc, PI
Gutermuth) were taken from the {\sl Herschel} \citep{pilbratt2010}
archive.  We selected these sources as examples because they cover
different ranges in distance, mass, and UV-field (Table~1). Dust optical depths
and temperatures were determined from a greybody fit to the surface
brightness at the 160 $\mu$m wavelength of PACS \citep{poglitsch2010}
together with the 250, 350, and 500 $\mu$m wavelengths of SPIRE
\citep{griffin2010}. For that, we assume a constant line-of-sight
temperature for each pixel. The optical depths are then converted into
H$_2$ column densities taking a dust opacity $\kappa_{\lambda}$ = 0.1
$\times$ ($\lambda$/300$\mu$m)$^{-\beta}$ cm$^2$/g with a fixed
$\beta$=2. Opacity variations were observed in Orion A \citep{roy2014}
so that we arrive to a relative accuracy of $\sim$50\% over the whole column
density range covered in {\sl Herschel} observations.  For more
details, we refer to \citet{hill2011}, \citet{schneider2015a}, and
\citet{vera2015}. For CepOB3, the PACS 160 $\mu$m data contain
nonfunctional bolometers.  The resulting column density maps thus show
a regular pattern with missing points, so that we fitted only the 250,
350, and 500 $\mu$m data points.
%We nevertheless compared the column
%density maps (and PDFs) obtained from the three- and four-wavelength
%fits and found that the differences in column density is less than
%$\sim$15\% and the PDF shape in terms of peak, deviation point and
%slopes does not change.
The angular resolution of the maps is 36$''$,
matched to the longest-wavelength {\sl Herschel} band.

%and we estimate their uncertainty to be 30--50\% (Russeil et al. 2013).

%The angular resolutions at 160 $\mu$m (PACS), 250 $\mu$m, 350 $\mu$m, and
%500 $\mu$m (all SPIRE), are $\sim$12$''$, $\sim$18$''$, $\sim$25$''$,
%and $\sim$36ß$''$, respectively. The SPIRE data were reduced with HIPE
%version 7.1956, including a destriper-module with a polynomial
%baseline of zeroth order.  Both scan directions were then combined using
%the `naive-mapper', i.e., a simple averaging algorithm.  The PACS data
%were reduced using HIPE 6.0.2106. In addition to the standard data
%reduction steps, non-linearity correction was applied on the 160
%$\mu$m signal, which affects only the bright ($>$ 10 Jy/pixel) regime.
%The level1 data were then combined into a map with Scanamorphos v10
%(Roussel \cite{roussel2012}).

\vspace{-0.6cm}
\section{Probablity distribution functions of GMCs} \label{pdfs}

Figure~\ref{pdf1} shows the column density maps and corresponding PDFs
for the GMCs of our study. To describe the PDF, we use the notation
p($\eta$) with
\begin{equation} p(\eta)d\eta=(2\,\pi\,\sigma^2_{\eta})^{-0.5} \,
  \exp[-(\eta-\mu)^2/(2\sigma^2_{\eta})]\,d\eta \end{equation} with
$\eta=\ln(N/\langle N \rangle)$, $\sigma_{\eta}$ as the
dimensionless dispersion of the logarithmic field, and $\mu$ as the mean.
%The normalization to the average column density $\langle N \rangle$
%allows for a direct comparison between clouds of different column
%density.
%, and the variance $\sigma_{\eta}$ is related to the density
%variations expected in a turbulent medium \citep{brunt2010}.
We determine $\sigma_{\eta}$ with a fit to the assumed lognormal low
column density part of the PDF and the slopes from a linear regression
power-law fit with $p(\eta) = p_0 (\eta/\eta_0)^s$ to the high-density
parts \citep[see][for more details]{schneider2015a,schneider2015b}.
%For low column densities, the PDFs are best fitted by a lognormal
%distribution. 
%\footnote{For the more distant clouds NGC6334 and W43, we
%  observe a more complex shape caused by line-of-sight contamination
%  but for simplicity, we keep a single lognormal fit.}.  
The intersection between lognormal and power-law tail regime lies at
extinctions\footnote{For better comparison to other studies in the
  literature, we use the visual extinction value derived from the
  column density adopting N(H$_2$)/\av=0.94$\times$10$^{21}$ cm$^{-2}$
  mag$^{-1}$ \citep{bohlin1978}.} \av\, of around 6--8 for CepOB3 and
MonR2 and \av$\sim$14 for NGC6334. Starting at these values, we then
fit a power-law to the distribution up to the break point where
additional high \av\ excess becomes apparent, i.e., at \av$\sim$40 for
MonR2, \av$\sim$60 for CepOB3, and \av$\sim$140 for NGC6334.  This
spread in values shows that the shift to a shallower power law does
not represent a universal threshold such as the one proposed for
massive star formation around \av = 300 \citep{krumholz2008}.  The
excess is clearly evident over a range of \av\ $\sim$40-200
for MonR2, \av\ $\sim$60-200 for CepOB3, and \av\ $\sim$140-300 for
NGC6334. Although it can be fitted by a power-law\footnote{The reduced
  $\chi^2_{red}$ values of the second power law fits scatter around
  unity (1.05$\pm$0.28, 1.08$\pm$0.10, 2.43$\pm$0.16 for CepOB3,
  MonR2, NGC6334) indicating that the general assumption of a
  power-law dependence is justified.}, we cannot exclude the
possibility that other distributions can fit the excess similarly well. In
considering its physical interpretation in comparison to the first
power-law tail, we keep the premise of a second power-law.

A word of caution is appropriate with regard to the PDF shape.
Additional extinction due to foreground/background emission provokes
that the width of the lognormal part becomes more narrow and the peak
shifts to higher \av, while the slope(s) of the power-law tail(s) are
steeper, compared to an undisturbed distribution
\citep{schneider2015a}.
However, the slope of the second power-law tail is less affected
because a contamination of a few \av\ does not significantly change
the highest column densities.  Contamination and projection
effects, however, could still lead to substructure in the PDF, which might complicate 
the simple lognormal + power-law tail(s) representation.
The accuracy of the lognormal fit is thus limited and the first
deviation point is not well defined.

\vspace{-0.6cm}
\section{Spatial identification of the different PDF regimes} \label{areas}

To identify the pixels that constitute the two power-law tails, we
outline in Fig.~\ref{pdf1} with a red and yellow contour the
respective \av-thresholds. While the red contour traces mainly
filaments, the yellow contour encloses individual {\sl clumps} with a
size scale of $\sim$1 pc, delimiting single (MonR2) or several
(CepOB3, NGC6334) star-forming center(s) of the GMCs. These central
regions correspond in their geometry and physical properties to what
was defined as 'ridges' (or hubs) in massive clouds
\citep{hill2011,hennemann2012}, i.e., regions with high column density
($N>$10$^{23}$ cm$^{-2}$) over small areas.  The PACS 70 $\mu$m images
(Fig.~\ref{pdf1}) indicate the stellar/protostellar activity in these
clumps: NGC6334 contains several \hii\ regions and a few ultra-compact
(UC) \hii\ regions \citep[e.g.,][]{russeil2013}, MonR2 hosts several UC
\hii\ regions and bipolar outflows \citep[e.g.,][]{fuente2010,dier2015},
and CepOB3 includes an embedded massive cluster.
%As a rough estimate
%of the total UV-field in the clouds, we give in Table~\ref{table} the
%average of the UV-field as well as the maximum value in the map (in a
%12$''$ beam). The field was derived following the method outlined in
%\citet{rocca2013}, i.e.,  assuming that all incident UV-radiation is
%re-emitted in the {\sl Herschel} 70 and 160 $\mu$m
%fluxes.
Table~\ref{table} lists the masses contained above the two break
points.  For the high column density regime (yellow), we determine
also the equivalent radius $r$, and the density $n$. It becomes
obvious that most of the mass is constituted by the lower column
density regime, i.e., 83\%, 61\%, and 68\% for CepOB3, MonR2, and
NGC6334, respectively. The mass above the first deviation point makes
up 15\% to 39\% of the complexes. The highest column density clumps
still account for a significant amout of mass in MonR2 and NGC6334
(both 15\%), but a much smaller proportion ($<$1\%) for CepOB3.  The
average density $\langle n \rangle$ in these clumps is also high,
i.e., between 2$\times$10$^4$ cm$^{-3}$ and 3.5$\times$10$^5$
cm$^{-3}$.  However, the existence of high density clumps/cores does
not by itself explain the second power law.  In a region such as
Aquila, $\sim$15\% of the total cloud mass resides in gas of density
larger than 10$^4$ cm$^{-3}$ \citep{vera2015}, but only a single
power-law tail is observed \citep{schneider2013,andre2014}. \\ We
focus now on the analysis of the two power-law tails.  From the slope
$s$, the exponent $\alpha$ for the density distribution
$\rho(r)\propto r^{-\alpha}$ can be derived. For a spherical geometry,
representing clumps and cores, $\alpha = -2/s +1$ \citep{fed2013}.
For a cylindrical one, characterizing filaments, $\alpha = -1/s +1$
\citep{myers2015}. Table~1 lists the values for both geometries for
the first and second tail.  The values of $\alpha$ for the first tail
range between 1.3 and 2, consistent with free-fall collapse, see
\citet{schneider2013,schneider2015a,schneider2015b} and
\citet{giri2014} for a detailed discussion of both filaments and
clumps/cores.  The values of $\alpha_{c2}$ between 1.85 and 2.64 
correspond very well to the range of values directly measured for the
B211/B213 filament in Taurus in \citet{pedro2013}.
Independent observational support for gravitational contraction comes
from molecular line observations of other GMCs that show spectral
infall signatures across ridges
\citep{schneider2010,schneider2015b,galvan2010,peretto2013}. These
studies show that the formation of high-mass stars requires a very
large mass accretion rate, provided by infall from merging filaments
and gravitational collapse of larger structures on parsec-scales.
For the excess -- presuming it can be described by a
power-law -- $\alpha$ lies between 1.9 and 4.2.  The assumption of a
spherical density distribution is probably more valid here, 
because the corresponding regions are more compact and circular.
However, free-fall alone can not produce $\alpha>$2, only a
slowed-down collapse can lead to piling up high (column) densities on
very small spatial scales.

\vspace{-0.6cm}
\section{Significance of the excess} \label{discuss}
 
The PDF excess we observe for the GMCs in our study is not a
systematic property of all GMCs. It is also seen in Rosette
\citep{schneider2012}, W43 and W51 (Schneider et al., in prep.), and W3
(Rivera-Ingraham et al., 2015), but there is no example of a low-mass star-forming
cloud that displays this feature. Similarly, there are also massive clouds
such as Carina and NGC3603 \citep{schneider2015a}, Vela
\citep{hill2011}, or M16 \citep{hill2012} where no clear indication
for this excess is found in the PDFs of the whole cloud.  Therefore,
it is unlikely that the excess is caused by a systematic bias
introduced during the determination of the dust column density maps.

It is out of the scope of this paper to discuss in detail the general
uncertainties related to the determination of the column density maps
and we refer the reader to
\citet{roy2013,roy2014} and \citet{vera2015}. Nevertheless, we mention here
briefly the biases that could potentially influence the
inferred slope and are thus relevant for the results of this paper: 1)
An overestimation of the line-of-sight temperature (assumed to be
constant for the SED fit) leads to an underestimation of the column
density (and the vice versa).  2) If dust opacity increases towards high
column densities, then this could give rise to a steeper slope of
the power-law tail at higher densities.  3) The adopted value of the
specific dust opacity $\beta$=2 could be too high in parts of the warm
GMCs. A lower (higher) value of $\beta$ in the greybody fits leads to
higher temperatures and thus lower (higher) column densities.

\vspace{-0.6cm}
\section{What causes the two power-law tails in the PDF?} \label{discuss}

Given that the excess/second power-law tail is only found in
high-density ($\langle n \rangle \sim$10$^{4-5}$ cm$^{-3}$)
star-forming clumps on size scales of $\sim$1 pc, i.e. in ridges/hubs,
the physical process(es) in control of it must be active on small
scales. It is evident that at these densities gravitational collapse
of individual star-forming cores is involved, but it is not yet understood what
produces the remarkable excess in column density.

We emphasize that in all cases the newly found excess is preceeded by
a first power-law in the PDF consistent with the dominant effect of
self-gravity found by numerical simulations
\citep[e.g.,][]{vaz2008,kritsuk2011} as well as observations
\citep[e.g.,][]{schneider2013,schneider2015a,schneider2015b}.  We thus
expect the regions under consideration to contract in almost free-fall
\citep{giri2014}, i.e., gas at a density $\rho_i$ \emph{falls} towards
higher densities $\rho_i'>\rho_i$. Without any change in physical
processes for higher densities, free-fall would proceed unimpeded and
the power-law would extend indefinitely.  Deviations from this
idealised single power-law can be twofold. On the one hand the second
excess could be due to a change in dynamics at high densities or due
to observational effects. In the following we give some tentative
explanations that, however, need more profound studies.

Any physical process that slows down the free-fall motions reduces the
flow of mass towards higher densities. We stress that in this paradigm
the deceleration of free-fall stems from \emph{within} the centres of
the cloud clumps. Processes acting from the outside like a hot ambient
medium in which the cloud is embedded or large-scale colliding flows
are unlikely to directly alter the PDF at the highest densities
without altering the distribution at lower densities that is in the
free-fall regime of the first power-law.  Rotational effects (1.) have
been invoked to explain excess at high densities seen in a density PDF
obtained from an isothermal, self-gravitating supersonic turbulence
simulation \citep{kritsuk2011}.  The spatial scales at which angular
momentum is likely to dominate, however, are probably smaller than the
resolution of the observations presented in this paper (0.1--0.25
pc). Likewise, thermodynamical effects of increasing thermal pressure
due to shielding and reduced cooling (2.)  is unlikely to dominate at
the observed scales \citep[e.g.,][]{larson2005}.  The role of magnetic
fields (3.)  is rather unclear. It was shown that {\sl strong}
\citep{koertgen2015} or {\sl intermediate} \citep{heitsch2001}
magnetic fields, acting on a clump scale, slow down and can even
completely prevent the star-formation process in magneto-hydrodynamic
(MHD) simulations. If magnetic fields entirely prevent the collapse
and star formation, the clouds are stable and will not show a first
gravitationally dominated power-law in the PDF. \emph{Moderate} and
\emph{weak} magnetic fields tend to reduce the degree of fragmentation
but once a gravitational instability sets in, the fields are unlikely
to stop further collapse and star formation \citep[e.g.,
][]{Ziegler2005, BanerjeePudritz2006}.
In this case, the N-PDF shows a power-law tail due to the
gravitational collapse of the supercritical cores.  Moreover, a change
in dominant geometry (4.), i.e., from filamentary to spherical, could
also provide an explanation. Longitudinal filament collapse on parsec
scales reduces the mass transfer rate and dense gas is then piled up  \citep{toala2012}.

All of the above proposed mechanisms to produce column density excess
should apply for all types of clouds -- not only to the most massive
GMCs. Therefore, high column density excesses should be observed more
commonly, but this is not the case. Note, however, that the slope of
the power-law tail(s) depends on the projection, i.e. the viewing
angle, in which the cloud is being observed \citep{ball2011}.  In any
case, a major difference between clouds forming only low-mass stars
and those with high-mass star-formation is stellar feedback (5.).  For
example, additional compression by {\sl internal} ionization due to an
ultra-compact \hii\ region can provoke a power-law tail with
$\alpha>2$ \citep{tremblin2014}.  (High-mass)
outflows may also play a role. It was shown by \citet{sadavoy2014}
that the slope of the power-law PDF in different regions in Perseus
depends on the local feedback from low-mass young stellar objects.  A
recent study of MonR2 \citep{dier2015} revealed a significant number
of CO outflows in the central region that imprints on the velocity
structure of the region.  Not all clouds with protostellar feedback,
however, also show a second power-law. If feedback does not
efficiently couple to the surroundings and the dense gas, an excess in
the PDF might not develop or might not be visible at the observed
scales in the column density PDF.

\vspace{-0.6cm}
\section{Conclusions and outlook}

At this point, it is not possible to give a final answer for our
detection of excess/second power-law in the PDF of massive
GMCs. Further studies are required to look into more detail which of
the proposed processes can play a dominant role. Besides more
sophisticated 3D hydrodynamic simualtions of star-forming regions a
deeper understanding of how local thermal and dynamical properties are
reflected in observations is needed.

\vspace{-0.6cm}
\section*{Acknowledgements}
%SPIRE has been developed by a consortium of institutes led by Cardiff
%Univ. (UK) and including: Univ. Lethbridge (Canada); NAOC (China);
%CEA, LAM (France); IFSI, Univ. Padua (Italy); IAC (Spain); Stockholm
%Observatory (Sweden); Imperial College London, RAL, UCL-MSSL, UKATC,
%Univ. Sussex (UK); and Caltech, JPL, NHSC, Univ. Colorado (USA). This
%development has been supported by national funding agencies: CSA
%(Canada); NAOC (China); CEA, CNES, CNRS (France); ASI (Italy); MCINN
%(Spain); SNSB (Sweden); STFC, UKSA (UK); and NASA (USA).  PACS has
%been developed by a consortium of institutes led by MPE (Germany) and
%including UVIE (Austria); KU Leuven, CSL, IMEC (Belgium); CEA, LAM
%(France); MPIA (Germany); INAF-IFSI/OAA/OAP/OAT, LENS, SISSA (Italy);
%IAC (Spain). This development has been supported by the funding
%agencies BMVIT (Austria), ESA-PRODEX (Belgium CEA/CNES (France), DLR
%(Germany), ASI/INAF (Italy), and CICYT/MCYT (Spain).
Part of this work was supported by the ANR-11-BS56-010 
``STARFICH'' and the ERC Grant 291294 ``ORISTARS''.
N.S. and V.O. acknowledge support by the DFG (Os 177/2-1 and 177/2-2) 
and central funds of the DFG-priority program 1573 (ISM-SPP).

%%%%%%%%%%%%%%%%%%%%%%%%%%%%%%%%%%%%%%%%%%%%%%%%%%



%\bibliographystyle{mnras}

\vspace{-0.6cm}
\begin{thebibliography}{99}
\bibitem[\protect\citeauthoryear{Andr\'e et al.}{2014}]{andre2014}
Andr\'e, Ph., et al., 2014, PPVI, Univ. of Arizon Press, 914, p.27
%Andr\'e, Ph.,  Di Francesco, J., Ward-Thompson, D., et al., 2014, PPVI, University of Arizon Press, 914, p.27
\bibitem[\protect\citeauthoryear{Ballesteros-Paredes et al.}{2011}]{ball2011}
Ballesteros-Paredes, et al., A., 2011, MNRAS, 416, 1436
%Ballesteros-Paredes, J., Vazquez-Semadeni, E., Gazol, A., 2011, MNRAS, 416, 1436
\bibitem[\protect\citeauthoryear{Banerjee et al.}{2006}]{BanerjeePudritz2006}
Banerjee, R., Pudritz, R., Anderson, D., 2006, MNRAS, 373, 1091
\bibitem[\protect\citeauthoryear{Bohlin et al.}{1978}]{bohlin1978} 
Bohlin, R.C., Savage, B.D., Drake, J.F., 1978, ApJ 224, 132
%\bibitem[\protect\citeauthoryear{Brunt et al.}{2010}]{brunt2010} 
%Brunt, C.M., Federrath, C., Price, D., 2010, MNRAS, 405, L56
%\bibitem[\protect\citeauthoryear{Brunt}{2015}]{brunt2015} 
%Brunt, C.M., 2015, MNRAS, 449, 4465  
%\bibitem[\protect\citeauthoryear{Butler et al.}{2014}]{butler2014}  
%Butler, M., Tan, J., Kainulainen, J., 2014,  ApJ, 782, L30
\bibitem[\protect\citeauthoryear{Dierickx et al.}{2015}]{dier2015} 
Dierickx, M., Jimenez-Serra, I., et al., 2015, ApJ, 803, 89
%Dierickx, M., Jimenez-Serra, I., Rivilla, V.M., Zhang, Q., 2015, ApJ, 803, 89
\bibitem[\protect\citeauthoryear{Didelon et al.}{2015}]{didelon2015} 
Didelon, P., Motte, F., Tremblin, P., et al., 2015, A\&A in press
\bibitem[\protect\citeauthoryear{Federrath \& Klessen}{2013}]{fed2013} 
Federrath, C., Klessen, R. S., 2013, ApJ, 763, 51 
\bibitem[\protect\citeauthoryear{Froebrich \& Rowles}{2010}]{froebrich2010} 
Froebrich, D., Rowles, J., 2010, MNRAS, 406, 1350
\bibitem[\protect\citeauthoryear{Fuente et al.}{2010}]{fuente2010} 
Fuente, A., Bern\'e, O., Cernicharo, P. et al., 2010, A\&A, 521, L23
\bibitem[\protect\citeauthoryear{Galvan-Madrid et al.}{2010}]{galvan2010}
Galvan-Madrid, R., Zhang, Q., et al., 2010, ApJ, 725, 17
%Galvan-Madrid, R., Zhang, Q., Keto, E., et al., 2010, ApJ, 725, 17
\bibitem[\protect\citeauthoryear{Girichidis et al.}{2014}]{giri2014}     
Girichidis, P., et al., 2014, ApJ, 781, 91    
%Girichidis, P., Konstandin, L., Whitworth, A.P., Klessen, R.S., et al., 2014, ApJ, 781, 91    
%\bibitem[\protect\citeauthoryear{Girichidis et al.}{2015}]{giri2015}     
%Girichidis, P., Schneider, N., Klessen, R.S., 2015, MNRAS, in prep.    
\bibitem[\protect\citeauthoryear{Griffin et al.}{2010}]{griffin2010}
Griffin, M., Abergel, A., Abreau, A., et al., 2010, A\&A, 518, L3 
\bibitem[\protect\citeauthoryear{Heitsch et al.}{2001}]{heitsch2001}
Heitsch, F., et al., 2001, ApJ, 547, 280
\bibitem[\protect\citeauthoryear{Hennemann et al.}{2012}]{hennemann2012}   
Hennemann, M., Motte, F., et al., 2012, A\&A, 543, L3  
\bibitem[\protect\citeauthoryear{Hill et al.}{2011}]{hill2011} 
Hill, T., Motte F., Didelon P., et al., 2011, A\&A, 533, 94 
\bibitem[\protect\citeauthoryear{Hill et al.}{2012}]{hill2012} 
Hill, T., Motte F., Didelon P., et al., 2012, A\&A, 542, 114
%\bibitem[1983]{hildebrand1983} 
%Hildebrand, R.H., 1983, QJRAS, 24, 267
\bibitem[\protect\citeauthoryear{Kainulainen et al.}{2009}]{kai2009}   
Kainulainen, J., Beuther, H., et al., 2009, A\&A, 508, L35   
%Kainulainen, J., Beuther, H., Henning, T., \& Plume, R., 2009, A\&A, 508, L35   
\bibitem[\protect\citeauthoryear{Kainulainen et al.}{2011}]{kai2011}   
Kainulainen, J., Beuther, H., et al., 2011, A\&A, 530, 64   
%Kainulainen, J., Beuther, H., Banerjee, R., et al., 2011, A\&A, 530, 64   
%\bibitem[\protect\citeauthoryear{Kainulainen et al.}{2013}]{kai2013}  
%Kainulainen, J., Tan, J.C., 2013, A\&A, 549, 53
\bibitem[\protect\citeauthoryear{Klessen}{2000}]{klessen2000}
Klessen, R.~S., 2000, ApJ, 535, 869 
\bibitem[\protect\citeauthoryear{K\"onyves et al.}{2015}]{vera2015}
K\"onyves, V., Andr\'e Ph., et al., 2015, A\&A, in press 
%K\"onyves, V., Andr\'e P., Men'shchikov A., et al., 2015, A\&A, submitted 
\bibitem[\protect\citeauthoryear{Koertgen et al.}{2015}]{koertgen2015}
Koertgen, B., Banerjee, R., 2015, A\&A sub., astro-ph: 1502.03306
\bibitem[\protect\citeauthoryear{Kritsuk et al.}{2011}]{kritsuk2011} 
Kritsuk, A.G., Norman, M.L., Wagner, R., 2011, ApJ, 727, L20
\bibitem[\protect\citeauthoryear{Krumholz \& McKee}{2008}]{krumholz2008} 
Krumholz, M., McKee, C., 2008, NATURE, 451, 1082
\bibitem[\protect\citeauthoryear{Larson}{2005}]{larson2005} 
Larson, R.B., 2005, MNRAS, 359, L211 
\bibitem[\protect\citeauthoryear{Motte et al.}{2010}]{motte2010} 
Motte, F., Zavagno A., Bontemps S., et al., 2010, A\&A, 518, L77
\bibitem[\protect\citeauthoryear{Myers}{2015}]{myers2015} 
Myers, P., 2015, ApJ, in press, astro-ph:1505.01124
\bibitem[\protect\citeauthoryear{Palmeirim et al.}{2013}]{pedro2013}
Palmeirim, P., Andr\'e, Ph., Kirk, J., et al., 2013, A\&A, 550, 38
\bibitem[\protect\citeauthoryear{Peretto et al.}{2013}]{peretto2013}   
Peretto, N.,  Fuller, G.A., et al., 2013, A\&A, 555, 112
%Peretto, N.,  Fuller, G.A., Duarte-Cabral, A., et al., 2013, A\&A, 555, 112
\bibitem[\protect\citeauthoryear{Pilbratt et al.}{2010}]{pilbratt2010}
Pilbratt, G., et al., 2010, A\&A 518, L1   
%Pilbratt, G., Riedinger, J., Passvogel, T., et al., 2010, A\&A 518, L1   
\bibitem[\protect\citeauthoryear{Poglitsch et al.}{2010}]{poglitsch2010}
Poglitsch, A., Waelkens, C., Geis, N., et al., 2010, A\&A 518, L2 
%\bibitem[\protect\citeauthoryear{Rayner et al.}{2015}]{rayner2015}
%Rayner, T., Griffin, M., Schneider, N., et al., 2015, in prep.
\bibitem[\protect\citeauthoryear{Rivera-Ingraham et al.}{2015}]{alana2015}   
Rivera-Ingraham, A., et al., 2015, ApJ, in press  
\bibitem[\protect\citeauthoryear{Roccatagliata et al.}{2013}]{rocca2013}   
Roccatagliata, V., Preibisch, T., et al., 2013, A\&A, 554, 6 
%Roccatagliata, V., Preibisch, T., Ratzka, T., Gaczkowski, B., 2013, A\&A, 554, 6 
\bibitem[\protect\citeauthoryear{Roy et al.}{2013}]{roy2013}   
Roy, A., Andr\'e, Ph., Palmeirim, P., et al., 2013, A\&A,    
\bibitem[\protect\citeauthoryear{Roy et al.}{2014}]{roy2014}   
Roy, A., Martin, P.G., Polychroni, D., et al., 2014, ApJ, 763, 55   
\bibitem[\protect\citeauthoryear{Russeil et al.}{2013}]{russeil2013}   
Russeil, D., Schneider, N., et al., 2013, A\&A, 554, 42 
%Russeil, D., Schneider, N., Anderson, L., et al., 2013, A\&A, 554, 42 
%\bibitem[2011]{price2011} 
%Price, D.J., Federrath, C., Brunt, C.M., 2011, ApJ, 727, L21
\bibitem[\protect\citeauthoryear{Sadavoy et al.}{2014}]{sadavoy2014} 
Sadavoy, S., Di Francesco, J., et al., 2014, ApJ, 787, L18
%Sadavoy, S., Di Francesco, J., Andr\'e, Ph., et al., 2014, ApJ, 787, L18
\bibitem[\protect\citeauthoryear{Schneider et al.}{2010}]{schneider2010} 
Schneider, N., Csengeri, T., et al., 2010, A\&A, 520, 49
%Schneider, N., Csengeri, T., Bontemps S., et al., 2010, A\&A, 520, 49
%\bibitem[2011]{schneider2011} 
%Schneider, N., Bontemps, S., Simon, R., et al., 2011, A\&A, 529, 1 
\bibitem[\protect\citeauthoryear{Schneider et al.}{2012}]{schneider2012} 
Schneider, N.,  Csengeri, T., et al., 2012, A\&A, 540, L11
%Schneider, N., Csengeri, T., Hennemann, M., et al., 2012, A\&A, 540, L11
\bibitem[\protect\citeauthoryear{Schneider et al.}{2013}]{schneider2013} 
Schneider, N., Andr\'e, P., K\"onyves, V., et al., 2013, ApJ, 766, L17 
\bibitem[\protect\citeauthoryear{Schneider et al.}{2015a}]{schneider2015a} 
Schneider, N., Ossenkopf, V., et al., 2015a, A\&A, 575, 79
%Schneider, N., Ossenkopf, V., Csengeri, T., et al., 2015a, A\&A, 575, 79
\bibitem[\protect\citeauthoryear{Schneider et al.}{2015b}]{schneider2015b} 
Schneider, N.,  Csengeri, T., et al., 2015b, A\&A, 578, 29
%Schneider, N., Csengeri, T., Klessen, R.S., et al., 2015b, A\&A, 578, 29
%\bibitem[\protect\citeauthoryear{Schneider et al.}{2015c}]{schneider2015c} 
%Schneider, N., Bontemps, S., Motte, F., et al., 2015c, A\&A, submitted
%\bibitem[\protect\citeauthoryear{Schneider et al.}{2015d}]{schneider2015d} 
%Schneider, Csengeri, C., Ossenkopf, V., et al., 2015d, A\&A, submitted
\bibitem[\protect\citeauthoryear{Toala et al.}{2012}]{toala2012} 
Toala, J.A., V\'azquez-Semadeni, E., et al., 2012, ApJ, 744, 190
%Toala, J.A., V\'azquez-Semadeni, E., Gomez, G.C., 2012, ApJ, 744, 190
\bibitem[\protect\citeauthoryear{Tremblin et al.}{2014}]{tremblin2014}    
Tremblin, P., Schneider, N., et al., 2014, A\&A, 564, 106
%Tremblin, P., Schneider, N., Minier, V., et al., 2014, A\&A, 564, 106
\bibitem[\protect\citeauthoryear{Vazquez-Semadeni et al.}{2008}]{vaz2008} 
Vazquez-Semadeni, E.,  et al., 2008, MNRAS, 390, 769
%Vazquez-Semadeni,E., Gonzales, R.F., Ballesteros-Paredes, J., et al., 2008, MNRAS, 390, 769
\bibitem[\protect\citeauthoryear{Ziegler}{2005}]{Ziegler2005} 
Ziegler, U., 2005, A\&A, 435, 385

\end{thebibliography}


%%%%%%%%%%%%%%%%%%%%%%%%%%%%%%%%%%%%%%%%%%%%%%%%%%

%%%%%%%%%%%%%%%%% APPENDICES %%%%%%%%%%%%%%%%%%%%%

%\appendix

%%%%%%%%%%%%%%%%%%%%%%%%%%%%%%%%%%%%%%%%%%%%%%%%%%

% Don't change these lines
\bsp	% typesetting comment
\label{lastpage}
\end{document}